\documentstyle[twoside,epsf]{article}

\catcode`\@=11
\long\def\@makefntext#1{
\protect\noindent \hbox to 3.2pt {\hskip-.9pt
$^{{\eightrm\@thefnmark}}$\hfil}#1\hfill}       

\def\@makefnmark{\hbox to 0pt{$^{\@thefnmark}$\hss}}    

\def\ps@myheadings{\let\@mkboth\@gobbletwo
\def\@oddhead{\hbox{}
\rightmark\hfil\eightrm\thepage}
\def\@oddfoot{}\def\@evenhead{\eightrm\thepage\hfil
\leftmark\hbox{}}\def\@evenfoot{}
\def\sectionmark##1{}\def\subsectionmark##1{}}

\oddsidemargin=\evensidemargin
\newcounter{sectionc}\newcounter{subsectionc}\newcounter{subsubsectionc}
\renewcommand{\section}[1] {\vspace{12pt}\addtocounter{sectionc}{1}
\setcounter{subsectionc}{0}\setcounter{subsubsectionc}{0}\noindent
    {\tenbf\thesectionc. #1}\par\vspace{5pt}}
\renewcommand{\subsection}[1] {\vspace{12pt}\addtocounter{subsectionc}{1}
\setcounter{subsubsectionc}{0}\noindent
{\bf\thesectionc.\thesubsectionc. {\kern1pt \bfit #1}}\par\vspace{5pt}}
\renewcommand{\subsubsection}[1] {\vspace{12pt}\addtocounter{subsubsectionc}{1}
    \noindent{\tenrm\thesectionc.\thesubsectionc.\thesubsubsectionc.
    {\kern1pt \tenit #1}}\par\vspace{5pt}}
\newcommand{\nonumsection}[1] {\vspace{12pt}\noindent{\tenbf #1}
    \par\vspace{5pt}}

\newcounter{appendixc}
\newcounter{subappendixc}[appendixc]
\newcounter{subsubappendixc}[subappendixc]
\renewcommand{\thesubappendixc}{\Alph{appendixc}.\arabic{subappendixc}}
\renewcommand{\thesubsubappendixc}
    {\Alph{appendixc}.\arabic{subappendixc}.\arabic{subsubappendixc}}

\renewcommand{\appendix}[1] {\vspace{12pt}
        \refstepcounter{appendixc}
        \setcounter{figure}{0}
        \setcounter{table}{0}
        \setcounter{lemma}{0}
        \setcounter{theorem}{0}
        \setcounter{corollary}{0}
        \setcounter{definition}{0}
        \setcounter{equation}{0}
        \renewcommand{\thefigure}{\Alph{appendixc}.\arabic{figure}}
        \renewcommand{\thetable}{\Alph{appendixc}.\arabic{table}}
        \renewcommand{\theappendixc}{\Alph{appendixc}}
        \renewcommand{\thelemma}{\Alph{appendixc}.\arabic{lemma}}
        \renewcommand{\thetheorem}{\Alph{appendixc}.\arabic{theorem}}
        \renewcommand{\thedefinition}{\Alph{appendixc}.\arabic{definition}}
        \renewcommand{\thecorollary}{\Alph{appendixc}.\arabic{corollary}}
        \renewcommand{\theequation}{\Alph{appendixc}.\arabic{equation}}
        \noindent{\tenbf Appendix \theappendixc #1}\par\vspace{5pt}}
\newcommand{\subappendix}[1] {\vspace{12pt}
        \refstepcounter{subappendixc}
        \noindent{\bf Appendix \thesubappendixc. {\kern1pt \bfit #1}}
    \par\vspace{5pt}}
\newcommand{\subsubappendix}[1] {\vspace{12pt}
        \refstepcounter{subsubappendixc}
        \noindent{\rm Appendix \thesubsubappendixc. {\kern1pt \tenit #1}}
    \par\vspace{5pt}}

\newcommand{\textlineskip}{\baselineskip=13pt}
\newcommand{\smalllineskip}{\baselineskip=10pt}

\newcommand{\copyrightheading}[1]
    {\vspace*{-2.5cm}\smalllineskip{\flushleft
    {\footnotesize Quantum Information and Computation, Vol.~1, No.~0 (2001) 000--000 #1}\\
    {\footnotesize \copyright\kern2pt Rinton Press}\\
     }}

\def\abstracts#1#2#3{{
    \centering{\begin{minipage}{4.5in}\footnotesize\baselineskip=10pt
    \parindent=0pt #1\par
    \parindent=15pt #2\par
    \parindent=15pt #3
    \end{minipage}}\par}}

\def\keywords#1{{
    \centering{\begin{minipage}{4.5in}\footnotesize\baselineskip=10pt
    {\footnotesize\it Keywords}\/: #1
     \end{minipage}}\par}}

\renewenvironment{thebibliography}[1]
        {\frenchspacing
     \ninerm\baselineskip=11pt
         \begin{list}{\arabic{enumi}.}
        {\usecounter{enumi}\setlength{\parsep}{0pt}
     \setlength{\leftmargin 12.7pt}{\rightmargin 0pt}
         \setlength{\itemsep}{0pt} \settowidth
    {\labelwidth}{#1.}\sloppy}}{\end{list}}

\newcounter{itemlistc}
\newcounter{romanlistc}
\newcounter{alphlistc}
\newcounter{arabiclistc}

\newcommand{\fcaption}[1]{
        \refstepcounter{figure}
        \setbox\@tempboxa = \hbox{\footnotesize Fig.~\thefigure. #1}
        \ifdim \wd\@tempboxa > 5in
           {\begin{center}
        \parbox{5in}{\footnotesize\smalllineskip Fig.~\thefigure. #1}
            \end{center}}
        \else
             {\begin{center}
             {\footnotesize Fig.~\thefigure. #1}
              \end{center}}
        \fi}

\newcommand{\tcaption}[1]{
        \refstepcounter{table}
        \setbox\@tempboxa = \hbox{\footnotesize Table~\thetable. #1}
        \ifdim \wd\@tempboxa > 5in
           {\begin{center}
        \parbox{5in}{\footnotesize\smalllineskip Table~\thetable. #1}
            \end{center}}
        \else
             {\begin{center}
             {\footnotesize Table~\thetable. #1}
              \end{center}}
        \fi}

\def\pmb#1{\setbox0=\hbox{#1}
    \kern-.025em\copy0\kern-\wd0
    \kern.05em\copy0\kern-\wd0
    \kern-.025em\raise.0433em\box0}

\def\fnt#1#2{\footnotetext{\kern-.3em
    {$^{\mbox{\scriptsize #1}}$}{#2}}}

\def\fpage#1{\begingroup
\voffset=.3in
\thispagestyle{empty}\begin{table}[b]\centerline{\footnotesize #1}
    \end{table}\endgroup}

\def\runninghead#1#2{\pagestyle{myheadings}
\markboth{{\protect\footnotesize\it{\quad #1}}\hfill}
{\hfill{\protect\footnotesize\it{#2\quad}}}}
\font\tenrm=cmr10
\font\tenit=cmti10
\font\tenbf=cmbx10
\font\bfit=cmbxti10 at 10pt
\font\ninerm=cmr9

\font\eightrm=cmr8

\def\FigName{figure}%
\newbox\captionbox
\long\def\@makecaption#1#2{%
  \ifx\FigName\@captype
    \vskip\abovecaptionskip
    \setbox\tempbox\hbox{{\figurecaptionfont #1\hskip1em #2}}
    \ifdim\wd\tempbox< 28pc
    \centerline{\box\tempbox}
    \else
    {\figurecaptionfont #1\hskip1em #2\par}
\fi\else
    \setbox\tempbox\hbox{{\tablecaptionfont #1\hskip1em #2}}
    \ifdim\wd\tempbox< 28pc
    \centerline{\box\tempbox}
    \else
    {\tablecaptionfont #1\hskip1em #2\par}%
    \fi
 \vskip\belowcaptionskip
 \fi}
\InputIfFileExists{psfig.sty}
{\typeout{^^Jpsfig.sty inputed...ok}}{\typeout{^^JWarning: psfig.sty could be be found.^^J}}
\InputIfFileExists{epsfsafe.tex}
{\typeout{^^Jepsfsafe.tex inputed...ok}}
            {\typeout{^^JWarning: epsfsafe.tex could not be found.^^J}}
\InputIfFileExists{epsfig.sty}
{\typeout{^^Jepsfig.sty inputed...ok}}{\typeout{^^JWarning: epsfig.sty could not be found.^^J}}
\InputIfFileExists{epsf.sty}
{\typeout{^^Jepsf.sty inputed...ok}}{\typeout{^^JWarning: epsf.sty could not be found.^^J}}%
\def\fps@figure{tbp}
\def\ftype@figure{1}
\def\ext@figure{lof}
\def\fnum@figure{Fig.\ \thefigure}
\def\qed{\hbox{${\vcenter{\vbox{              
   \hrule height 0.4pt\hbox{\vrule width 0.4pt height 6pt
   \kern5pt\vrule width 0.4pt}\hrule height 0.4pt}}}$}}

\newcommand{\DE}[1]{\left\{ #1 \right\}}

\newcommand{\ket}[1]{\left| #1 \right\rangle}
\newcommand{\bra}[1]{\left\langle #1 \right|}
\newcommand{\braket}[2]{\left\langle #1 \mid #2 \right\rangle}

\newcommand{\op}[1]{{\mathbf{#1}}}

\newcommand{\tr}{\mathrm{Tr}}

\newcommand{\eg}{{\it{e.g.}}: }
\newcommand{\ie}{{\it{i.e.}}: }
\newcommand{\etal}{{\it{et al.}} }

\newcommand{\pr}[4]{(#1), {\it{#2}} {Phys. Rev.,} {#3,} {#4}}
\newcommand{\pra}[4]{(#1), {\it{#2}} {Phys. Rev. A,} {#3,} {#4}}

\newcommand{\prl}[4]{(#1), {\it{#2}} {Phys. Rev. Lett.,} {#3,} {#4}}

\newcommand{\pla}[4]{(#1), {\it{#2}} {Phys. Lett. A,} {#3,} {#4}}

\newcommand{\nat}[4]{(#1), {\it{#2}} {Nature,} {#3,} {#4}}

\newcommand{\pt}[4]{(#1), {\it{#2}} {Phys. Today,} {#3,} {#4}}

\begin{document}
\setlength{\textheight}{8.0truein}    

\runninghead{How to fool the GHZ and W witnesses   $\ldots$}
            {D. Cavalcanti and M.O. Terra Cunha $\ldots$}

\normalsize\textlineskip
\thispagestyle{empty}
\setcounter{page}{1}


\vspace*{0.88truein}

\fpage{1} \centerline{\bf How to fool the GHZ and W witnesses}
\vspace*{0.035truein} \centerline{\bf} \vspace*{0.37truein}
\centerline{\footnotesize D. Cavalcanti
\footnote{dcs@fisica.ufmg.br}} \vspace*{0.015truein}
\centerline{\footnotesize\it Departamento de F\'{\i}sica,
Universidade Federal de Minas Gerais}\baselineskip=10pt
\centerline{\footnotesize\it Belo Horizonte, Minas Gerais, CP 702,
30123-970, Brazil} \vspace*{10pt} \centerline{\footnotesize M.O.
Terra Cunha \footnote{tcunha@mat.ufmg.br}} \vspace*{0.015truein}
\centerline{\footnotesize\it Departamento de Matem\'atica,
Universidade Federal de Minas Gerais}\baselineskip=10pt
\centerline{\footnotesize\it Belo Horizonte, Minas Gerais, CP 702,
30123-970, Brazil} 

\vspace*{0.21truein} \abstracts{In spite of the fact that there are
only two classes of three-qubit genuine entanglement: W and GHZ, we
show that there are three-qubit genuinely entangled states which can
not be detected neither by W nor by GHZ entanglement witnesses.}{}{}

\vspace*{10pt} \keywords{entanglement witness, detection of
entanglement, tripartite entanglement}

\vspace*{1pt}\textlineskip  

\vspace*{-0.5pt} \noindent

\section{Introduction}
Entanglement is presumably the most non-classical feature of quantum
mechanics. It is in the heart of Einstein-Podolsky-Rosen paradox
\cite{EPR}; of the so-called {\emph{which way}} interferometers
\cite{WW}, in which the interference pattern is washed out by the
entanglement of the interferometric particle with a ``which way
discriminator''; of teleportation of quantum states \cite{tele}; of
some cryptographic protocols \cite{Eke92}; and of the most important
quantum algorithms \cite{alg}.  It is not difficult to
{\emph{define}} when a quantum state is entangled:  for pure states
they can only be (in bipartite case) factorizable,  when can be
written as $\ket{\psi}\otimes \ket{\phi}$, or entangled.   The
multipartite case is a little bit subtler, since the parts can be
put together. For example, for three qubits ($A$, $B$, and $C$), the
state given by a Bell state \cite{Bell}   $\ket{\Psi ^{\pm}} =
\DE{\ket{01} \pm \ket{10}}/\sqrt{2}$ of the pair $AB$ together with
a pure state of $C$ is said {\emph{biseparable}}, in the sense that,
with respect    to the bipartition $AB - C$, it has no entanglement
at all. One usually says that in such a state there is no
{\emph{genuine}} tripartite entanglement \cite{Poligamia}. For mixed
states, whenever $\op{\rho}$     can be written as an ensemble of
pure states which do not have some kind of entanglement,     the
mixed state is also said not to have such kind of entanglement.
However, it is not a simple task to identify whether a given state
has some kind of entanglement. Peres obtained a good criterion by
noting that for separable states, the {\emph{partial transposition}}
lead to valid density matrices, whereas for some entangled state a
non-positive matrix is generated \cite{Per}. The Horodeckis put this
contribution in the context of  {\emph{positive maps which fail to
be completely positive}} acting on density operators \cite{Hor}, and
showed that the criterion is sufficient only for two qubits or for
one qubit and one qutrit. Examples of bipartite entangled states
which have positive partial transpose (PPT) are known \cite{PPT}.

Another way of identifying entanglement comes from the fact that,
for a given type of entanglement, the set of (mixed) states which
are free from such entanglement is a closed convex set. Any point in
the complement of a closed convex subset of a vector space can be
separated from it by a hyperplane \cite{Roc}. Equivalently, there is
a linear functional over this space which is positive in all
separable states $\sigma$ (\ie those which do not pursue the
inspected kind of entanglement), and negative in the specific
entangled state $\rho$ (\ie the point out of the closed convex
subset). This functional is called an {\emph{entanglement witness}},
and is represented by the operator ${\mathcal{W}}$ \cite{Terhal}. In
linear algebraic context, it reads: $\tr{({{\mathcal{W}}}\rho)}<0$
and $\tr{({\mathcal{W}}\sigma)}\geq0$, for all separable $\sigma$.
Once again, it is simple {\emph{in theory}}, but given a state which
one wants to decide whether it shows some kind of entanglement or
not, it is not easy to find the appropriated witness. A very
important class of entanglement witnesses (EW) is the optimal
entanglement witnesses (OEW). An OEW is the EW that best witnesses
the entanglement of $\rho$, in the sense that it reaches the maximum
value of $\mid \tr{({\mathcal{W}}\rho)}\mid$ \cite{cirac,barb}.
Unfortunately, finding  an OEW is a difficult task because it
involves a hard process of optimization \cite{NPhard}. However some
improvements has been done in the sense of finding OEWs for certain
kinds of states \cite{Toth,?}.

For three qubits, D\"ur, Vidal, and Cirac showed that there are two
different genuine tripartite entangled states, in the sense that,
even in a statistical sense, they can not be locally converted one
into another \cite{Ciracetal}. These states are the so called GHZ
state, which appeared on the literature in Ref. \cite{GHZ},
$\ket{GHZ} = \DE{\ket{000} + \ket{111}}/\sqrt{2}$, and the W state
\cite{W}, $\ket{W} = \DE{\ket{001} + \ket{010}
+\ket{100}}/\sqrt{3}$. Ishizaka and Plenio showed that even with
PPT-operations (those which send PPT operators into PPT operators),
GHZ and W remain inequivalent \cite{IP05}, however, stochastically
there is a protocol to convert GHZ in W via PPT-operations with
approximately $75\%$ of success \cite{IP04}.

\section{Fooling the GHZ/W-criterion}
A systematic  way to construct entanglement witnesses to pure
states is looking for the optimal element of the set of EWs that
have the specific form \cite{detect3}
\begin{equation}
{{\mathcal{W}}}_{\psi} = \Lambda - \ket{\psi}\bra{\psi},
\end{equation}
with $\Lambda \in {\mathcal{R}}$ and multiplication by the identity
operator omitted in notation. By the condition that
$\tr{(W\sigma)}\geq0$ if $\sigma$ is separable (denote $\mathcal{S}$
this set), we can see that
\begin{equation}
\Lambda=\max_{\sigma \in {\mathcal{S}}} \|
\braket{\sigma}{\psi}\|^2.
\end{equation}
As an example, for the specific case of three qubits, the OEW for
the GHZ-like states,
\begin{equation}
\ket{GHZ(\phi)}=\frac{(\ket{000}+e^{i\phi}\ket{111})}{\sqrt{2}}
\end{equation}
and for the W-like states,
\begin{equation}
\ket{W(\gamma ,
\beta)}=\frac{(\ket{001}+e^{i\gamma}\ket{010}+e^{i\beta}\ket{100})}{\sqrt{3}},
\end{equation}
are
\begin{equation}
{{\mathcal{W}}}_{GHZ(\phi)}=\frac{1}{2}-\ket{GHZ(\phi)}\bra{GHZ(\phi)}
\end{equation}
 and
 \begin{equation}
 {\mathcal{W}}_{W(\gamma , \beta)}=\frac{2}{3}-\ket{W(\gamma ,\beta)}\bra{W(\gamma , \beta)},
 \end{equation}
 respectively. Motivated by the existence of the two classes of
genuine three qubit entanglement, some authors have used the
GHZ-witness and the W-witness as a test of existence of genuine
tripartite entanglement. We call this test the GHZ/W-criterion
and, as we shall see, this criterion is incomplete in the sense
that there are genuine entangled states which are not detected by
it.

In the context of generalizing the Schmidt decomposition, Ac\'in
\etal showed that all three-qubit pure states can be written in the
following way \cite{acin}:
\begin{equation}
\ket{\psi}=\lambda_0
\ket{000}+\lambda_1e^{i\alpha}\ket{001}+\lambda_2
\ket{010}+\lambda_3\ket{100}+\lambda_4 \ket{111},
\end{equation}
where $0 \leq \lambda_i \in \mathcal{R}$, $0\leq \alpha \leq \pi$,
and $\sum_i \lambda_{i}^2=1$, by appropriate choices of local basis.
Let us see what are the conditions for the GHZ$(\phi)$-witness and
the W$(\gamma, \beta)$-witness indicating entanglement in a general
pure state.

GHZ$(\phi)$-witness:
\begin{equation}
\tr({\mathcal{W}}_{GHZ(\phi)}\ket{\psi}\bra{\psi})=\frac{1}{2}-\frac{1}{2}
(\lambda_{0}^2+\lambda_{4}^2+2\lambda_{0}\lambda_{4}\cos(\phi-\alpha))<0.
\end{equation}
As $\lambda _0, \lambda _4 \geq 0$, it is sufficient to our aim to
consider $\phi=\alpha$. Thus,
\begin{equation}(\lambda_0 + \lambda_4)^2>1,
\end{equation}
must hold for the entanglement of $\ket{\psi}$ to be detected by
GHZ$(\phi)$-witness.

W$(\gamma, \beta)$-witness:
\begin{eqnarray}
\tr({\mathcal{W}}_{W(\gamma,\beta)}\ket{\psi}\bra{\psi})&=&\frac{2}{3}-
\frac{1}{3}\{\lambda_{1}^2+\lambda_{2}^2+\lambda_{3}^2+2\lambda_{1}\lambda_{2}\cos(\gamma-\phi)+\nonumber\\
&&2\lambda_{1}\lambda_{3}\cos(\beta-\phi)+2\lambda_{2}\lambda_{3}\cos(\gamma-\beta)\}<0.
\end{eqnarray}
Again, we consider the extremal case $\gamma=\phi=\beta$. Thus,
\begin{equation}
(\lambda_1+\lambda_2+\lambda_3)^2>2.
\end{equation}

\begin{figure} [htbp]
\vspace*{13pt}
\centerline{\psfig{file=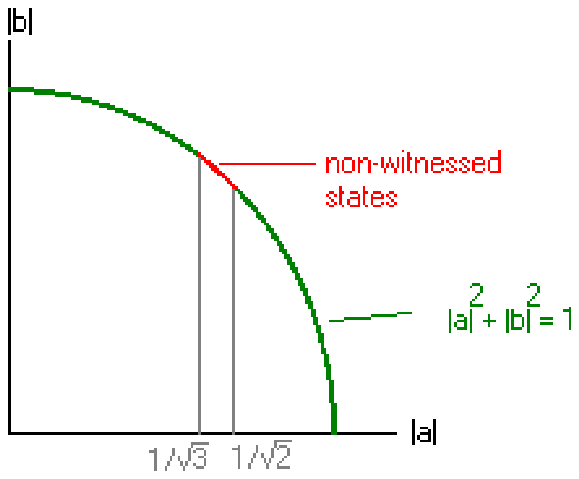, width=8.2cm}} 
\vspace*{13pt} \fcaption{\label{fig:ab} Red: family of states
$\ket{\xi '}=a\ket{GHZ(\phi)}+b\ket{W(\gamma,\beta)}$ which are
not detected by the GHZ/W-criterion.}
\end{figure}

Within these conditions, it is easy to see that the state
\begin{equation}
\ket{\xi}=\frac{1}{\sqrt{5}}(\ket{000}+\ket{001}+\ket{010}+\ket{100}+\ket{111})
\end{equation}
is neither witnessed by W$(\gamma, \beta)$-witness nor by
GHZ$(\phi)$-witness. Note that $\ket{\xi}$ has genuine-tripartite
entanglement, \ie it cannot be written as a  biseparable state. In
fact, $\ket{\xi}$ is a particular case of a whole family of pure
states which are not witnessed by ${\mathcal{W}}_{GHZ(\phi)}$ and
${\mathcal{W}}_{W(\gamma,\beta)}$. The states
\begin{equation}
\ket{\xi '}=a\ket{GHZ(\phi)}+b\ket{W(\gamma,\beta)},
\end{equation}
with $|a|^2 +|b|^2 =1$, reach
\begin{equation}
\langle {\mathcal{W}}_{GHZ(\phi)} \rangle =\frac{1}{2}-|a|^2,
\end{equation}
and
\begin{equation}
\langle {\mathcal{W}}_{W(\gamma, \beta)} \rangle =\frac{2}{3}-|b|^2.
\end{equation}
It means that $\ket{\xi '}$ is witnessed if $|a|^2>\frac{1}{2}$ or
$|b|^2>\frac{2}{3}$. However these conditions exclude all states
which satisfy $ \frac{1}{3} \leq |a|^2 \leq \frac{1}{2}$ (see
figure \ref{fig:ab}).

Furthermore statistic mixtures of unwitnessed pure states are also
unwitnessed. To prove it we can take a look at the state:
\begin{equation}
\rho=\sum_i p_i \ket{\xi'_i}\bra{\xi'_i},
\end{equation}
where the states $\ket{\xi'_i}$ are not witnessed by the
GHZ/W-criterion. Applying the criterion to $\rho$ we have:
\begin{equation}
\tr({\mathcal{W}}_{GHZ(\phi)}\rho)=\sum_{i}p_i\tr({\mathcal{W}}_{GHZ(\phi)}\ket{\xi'_i}\bra{\xi'_i})\geq0
\end{equation}
and
\begin{equation}
 \tr({\mathcal{W}}_{W(\gamma, \beta)}\rho)=\sum_{i}p_i\tr({\mathcal{W}}_{W(\gamma,\beta)}\ket{\xi'_i}\bra{\xi'_i})\geq0,
\end{equation}
where we have used the linearity of trace and that $p_i\geq0$,
$\tr({\mathcal{W}}_{GHZ(\phi)}\ket{\xi'_i}\bra{\xi'_i})\geq0$, and
$\tr({\mathcal{W}}_{W(\gamma,\beta)}\ket{\xi'_i}\bra{\xi'_i})\geq0$.
For sure, it is harder to guarantee which such combinations remain
tripartite genuine entangled, however, continuity is enough to
guarantee that some of them do remain.

\section{Conclusion and Speculations}
In conclusion, it was shown that testing the GHZ$(\phi)$-witnesses
and the W$(\gamma, \beta)$-witnesses is not a sufficient condition
to attest if a general state is a genuine-entangled state. We have
presented a family of pure states which fools the GHZ/W-criterion.
For sure, this must not be a great surprise, since entanglement
witness are taylored to indicate entanglement, not to deny it. We
believe that our work can avoid mistakes on the detection of
three-qubit entanglement in future works.

An interesting question raised by this incomplete criterion is to
determine whether there is a finite set of witness operators that
can determine if any state is a genuine-entangled state. Maybe it
could be done through recent results on searching procedures for OEW
\cite{NPhard,Ferpra}. It can be viewed as a detection counterpart of
the problem of determining the minimal set of entangled states that
can generate all entangled states under certain operations (\eg
LOCC) \cite{IP05}. It could help, for instance, in understanding the
geometry behind the set of separable states with respect to each
kind of entanglement, in a complementary way to Ref. \cite{Lev05}.
Another elucidating research would consist in finding what are the
states witnessed by the class of witnesses reached through of
different local basis of the GHZ and W states.

\nonumsection{Acknowledgements} \noindent
 D. Cavalcanti acknowledges financial support of the Brazilian agency CNPq. The
authors thank Fernando Brand\~ao for stimulating discussions on
the theme.

\nonumsection{References} 

\end{document}